%% file: EdgeControl_arxiv.tex
\documentclass[%
 aps,prl,twocolumn,superscriptaddress,floatfix
]{revtex4-2}

\usepackage{graphicx}% Include figure files
\usepackage{dcolumn}% Align table columns on decimal point
\usepackage{bm}% bold math
\usepackage[utf8]{inputenc}
\usepackage[T1]{fontenc}
\usepackage{etoolbox}
\usepackage{newtxtext}
\usepackage[varvw]{newtxmath}

\makeatletter
\def\@email#1#2{%
 \endgroup
 \patchcmd{\titleblock@produce}
  {\frontmatter@RRAPformat}
  {\frontmatter@RRAPformat{\produce@RRAP{*#1\href{mailto:#2}{#2}}}\frontmatter@RRAPformat}
  {}{}
}%
\makeatother

\begin{document}

\title{Spatiotemporal imaging of gate-controlled multipath dynamics\\
of fractional quantum Hall edge excitations}
% Force line breaks with \\

\author{Yunhyeon Jeong}
\affiliation{Department of Physics, Tohoku University, Sendai 980-8578, Japan}

\author{Akinori Kamiyama}
\affiliation{Department of Physics, Tohoku University, Sendai 980-8578, Japan}

\author{John N. Moore}
\affiliation{Department of Physics, Tohoku University, Sendai 980-8578, Japan}

\author{Takaaki Mano}
\affiliation{National Institute for Materials Science, Tsukuba, Ibaraki 305-0047, Japan}

\author{Yuuki Sugiyama}
\affiliation{Institute for Solid State Physics, The University of Tokyo, Chiba, 277-8581, Japan}

\author{Tokiro Numasawa}
\affiliation{Institute for Solid State Physics, The University of Tokyo, Chiba, 277-8581, Japan}

\author{Ken-ichi Sasaki}
\affiliation{NTT Research Center for Theoretical Quantum Information, NTT, Inc., 3-1 Morinosato Wakamiya, Atsugi, Kanagawa 243-0198, Japan}

\author{Masahiro Hotta}
\affiliation{Department of Physics, Tohoku University, Sendai 980-8578, Japan}

\author{Go Yusa}
\affiliation{Department of Physics, Tohoku University, Sendai 980-8578, Japan}

\date{\today}

%TC:ignore
\begin{abstract}
We image gate-controlled multipath dynamics of edge excitations in a $\nu=1/3$ fractional quantum Hall device using stroboscopic photoluminescence (PL) microscopy and spectroscopy. A control gate switches the response between mesa- and gate-defined trajectories and reveals a regime where both appear at the gated-area entrance. PL spectra measured far downstream along the mesa edge show that the excitation arrives after passing through the intermediate region between these trajectories. This demonstrates a gate-tuned propagation geometry rather than motion along a single well-defined boundary path.

\end{abstract}

%TC:endignore
\maketitle

Recent advances in controllable condensed-matter and optical platforms have broadened the scope of analog experiments beyond their original astrophysical motivation \cite{unruh1981experimental,garay2000sonic,schutzhold2002gravity,Wilson2011,weinfurtner2011measurement,Sakharov_Osc,Steinhauer2016,MunozdeNova2019,tani2024violation,Liou2019}. Among candidate platforms, quantum Hall systems are particularly attractive because their chiral edge excitations admit low-energy effective descriptions closely related to conformal field theories (CFTs) \cite{HottaPRA14,Stone_2013,Hegde,NambuPRD23,YOSHIMOTO2025130100,sugiyama2025anomalous}. This edge-based description provides a route to analog-spacetime proposals, in which the edge serves as a one-dimensional analogue spacetime, the propagating edge magnetoplasmon plays the role of light, and a designed electrostatic potential landscape defines the effective geometry through which the excitation propagates \cite{HottaPRD22}.

Real devices, however, depart substantially from idealized one-dimensional edge descriptions. The electrostatic landscape experienced by an edge excitation contains not only intentional gate-defined potentials but also built-in potential variations arising from the heterostructure, surface, and device fabrication \cite{spicer1979new,freeouf1981schottky,thornton1986one,simmons1988quantum,nieder1990one,ilani2004microscopic,martin2004localization,hayakawa2013real}. Such a spatially varying landscape can redirect edge excitations, modify their propagation time, and change their spatial profile, thereby determining the effective propagation geometry actually realized in the device. This sensitivity is essential for analog-spacetime experiments, but it also means that the realized geometry may differ substantially from an idealized one-dimensional path.

Under static conditions, local probes such as scanning single-electron transistors and micro-photoluminescence (PL) have provided valuable insight into the bulk fractional quantum Hall state by revealing inhomogeneity in the local electronic landscape \cite{ilani2004microscopic,martin2004localization,hayakawa2013real}. These measurements establish that the electrostatic landscape in a fractional quantum Hall device can be spatially nonuniform. They do not, however, directly address how a launched edge excitation propagates through such a landscape.

In addition, spatial variation of the electrostatic potential energy $U$ across a local edge can make the propagation speed nonuniform. As a qualitative guide, with $y$ chosen locally as the coordinate normal to the edge, the $E\times B$ drift velocity scale is set by $(eB)^{-1}dU/dy$, suggesting that changes in the steepness of the confinement can be reflected in the propagation time \cite{yoshioka2013quantum}.

\begin{figure*}
    \centering
    \includegraphics{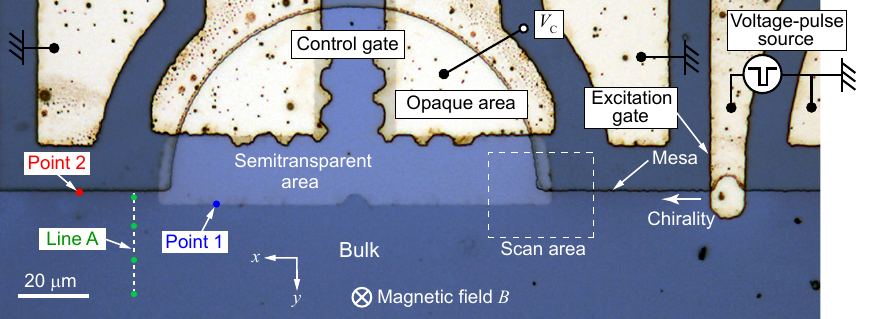}    
    \caption{Micrograph showing the device layout for gate-controlled propagation of fractional quantum Hall edge excitations. The excitation gate, driven by a voltage-pulse source, is located 42~$\mu$m upstream of the control-gated region along the mesa edge. The semi-circular control gate consists of an opaque Ti/Au region ($\sim$300~nm thick) and a semi-transparent Au region ($\sim$10~nm thick) with an optical transmittance of $50\%$ \cite{eytan1998near,yusa2000onset}. The chirality is set by the magnetic field $B$, and the launched excitation propagates from right to left along the mesa edge. The dashed rectangle indicates the scan area used for Fig.~\ref{fig:mappings}. Points~1 and~2 denote the fixed measurement positions used in Fig.~\ref{fig:time-resolve}. Line~A indicates the measurement path used for Fig.~\ref{fig:microPLspectra}. The grounded electrodes form part of the coplanar waveguide structures \cite{france2025electrically}.}%
\label{fig:device}
\end{figure*}

Here we report spatiotemporal imaging of gate-controlled multipath dynamics of edge excitations in a $\nu=1/3$ fractional quantum Hall device using stroboscopic time-resolved micro-photoluminescence microscopy and spectroscopy with $\sim$100-ps resolution \cite{kamiyamaPRR,kamiyamaAPL,france2025electrically,jeong2026direct}. By tuning a control-gate-defined potential landscape, we observe switching between mesa-defined and gate-defined trajectories and identify an intermediate regime in which the excitation-induced response appears on both trajectories, indicating that the response is not pinned to a single well-defined boundary path. PL spectra measured far downstream along the mesa edge further show that the excitation arrives after passing through the intermediate region between the mesa- and gate-defined trajectories. These observations, together with the gate-dependent time traces and temporal broadening, show that propagation is strongly reshaped by the real electrostatic landscape.

The sample was prepared from a GaAs/Al$_{0.2}$Ga$_{0.8}$As heterostructure containing a 15-nm quantum well (QW) \cite{kamiyamaPRR}. A Si-doped substrate was used as a back gate to control the electron density $n$ in the QW. As shown in Fig.~\ref{fig:device}, the device consists of a semi-circular mesa with a radius of $50~\mu$m connected to straight mesa sections. A semi-circular control gate covers the corresponding mesa region, while an excitation gate is placed upstream of the control-gated section with respect to the chirality set by $B$. A voltage pulse applied to the excitation gate launches an edge excitation that first propagates along the upstream mesa edge, enters the control-gate region, and then exits into the downstream ungated mesa edge. Additional experimental details are provided in End Matter (Appendix A). Unless otherwise noted, the measurements were performed in the $\nu=1/3$ fractional quantum Hall state at $B=14$~T and $40$--$50$~mK. 

We use stroboscopic PL spectroscopy as a local optical probe of the time-dependent electronic environment near the edge. Optical pulses with a width of $\sim 1$~ps and a repetition period of $\sim 13$~ns are synchronized with the voltage pulse applied to the excitation gate, allowing the PL response to be sampled at a controlled delay after each electrical excitation. The optical excitation creates charged excitons (trions), whose PL spectrum near $\nu=1/3$ contains singlet and triplet features associated with different electron-spin configurations \cite{wojs2000charged,yusaPRL01,hayakawa2013real}. Although the microscopic interpretation of trion spectra in the fractional quantum Hall regime is nontrivial, changes in the trion PL spectrum provide a sensitive local readout of the surrounding two-dimensional electron system \cite{yusaPRL01,hayakawa2013real,france2025electrically}. The delay $t$ is defined for each measurement configuration by choosing a delay origin, as described in the Supplemental Material (SM).

\begin{figure}
    \includegraphics[scale=1.0, pagebox=artbox, clip]{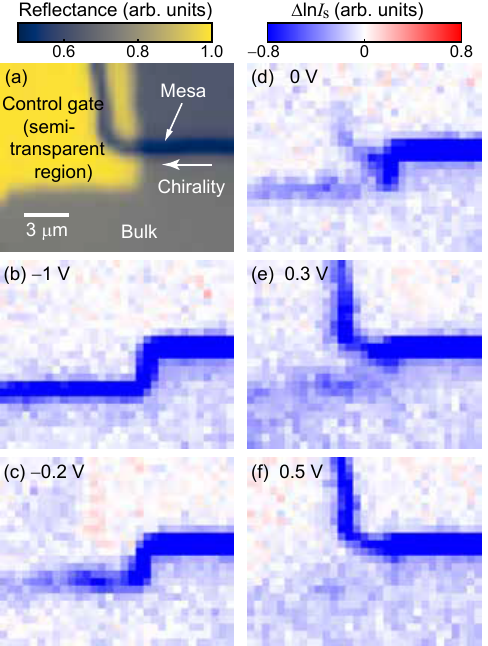}
 
    \centering
    \caption{Gate-voltage-dependent spatial response of an edge excitation measured over a $45 \times 24~\mu\mathrm{m}^2$ scan area at the entrance to the control-gate region. (a) Reflectance map measured with the same optical setup at $40$--$50$~mK; the yellow region corresponds to the semi-transparent part of the control gate. (b)--(f) Time-resolved snapshots of the excitation-induced logarithmic change in the singlet PL intensity, 
    $\Delta \ln I_{\mathrm{s}}$,
    as defined in the main text.
    Negative values indicate suppression of the singlet PL by the edge excitation, whereas values near zero indicate little or no change. The rectangular pulse amplitude applied to the excitation gate is $\pm 100$~mV. The control-gate voltages are (b) $V_\mathrm{c}=-1.0$~V, (c) $-0.2$~V, (d) $0$~V, (e) $0.3$~V, and (f) $0.5$~V.}%
    \label{fig:mappings}%
\end{figure}

First, to investigate microscopic control of the edge-propagation pathway, we focused on the region where the upstream excitation enters the control-gate area (the scan area in Fig.~\ref{fig:device}). We acquired stroboscopic PL spectra over this scan area while varying the DC voltage $V_\mathrm{c}$ applied to the control gate (Fig.~\ref{fig:mappings}). To characterize the propagation pathway, we analyzed time-resolved snapshots of the excitation-induced logarithmic change in the singlet PL intensity, $\Delta \ln I_\mathrm{s}\equiv \ln[I_\mathrm{s}/I_\mathrm{s}^0]$,  
at each spatial point (exposure time $20$~s); see the SM for details. Here, $I_\mathrm{s}$ 
is the integrated singlet PL intensity measured at $t = 0.6$~ns, and $I_\mathrm{s}^0$  
is the reference value measured at $t = -1.9$~ns, before the excitation reaches the focal point. Negative values of $\Delta \ln I_\mathrm{s}$ 
(blue regions) indicate suppression of the integrated singlet PL intensity by the edge excitation \cite{kamiyamaPRR,kamiyamaAPL,france2025electrically}. The systematic evolution of these patterns with $V_\mathrm{c}$ provides direct real-space evidence for gate-controlled redirection of the edge-excitation pathway.

This behavior is consistent with the picture that edge channels form where the local potential crosses the Fermi energy $E_\mathrm{F}$. For sufficiently negative $V_\mathrm{c}$ (e.g., $V_\mathrm{c}=-1$~V in Fig.~\ref{fig:mappings}(b)), the potential beneath the control gate is raised such that the edge forms near the boundary between the gate-covered region and the ungated GaAs surface (the gate boundary). Edge excitations then propagate along this boundary, hereafter referred to as the gate-defined path (Fig.~\ref{fig:mappings}(b)). In contrast, for positive $V_\mathrm{c}$ (e.g., $V_\mathrm{c}=0.5$~V in Fig.~\ref{fig:mappings}(f)), the potential beneath the gate is lowered and the edge forms near the mesa boundary, so excitations follow the mesa-defined path (Fig.~\ref{fig:mappings}(f)).

Notably, access to both propagation paths is observed over a relatively broad range of $V_\mathrm{c}$, from $0$ to $0.3$~V. Using a simple capacitive coupling model~\cite{france2025electrically}, this corresponds to an estimated electrochemical-potential shift of $\sim 4$~meV beneath the control gate, comparable to the $\sim 3$~meV shift expected directly under the excitation gate during the $0.2$~V$_{\mathrm{pp}}$ (i.e., $\pm 0.1$~V) pulse. This comparison suggests that, in this intermediate voltage range, the measured response is not pinned to a single well-defined boundary path. Instead, the static potential landscape set by $V_\mathrm{c}$ and the transient excitation-gate perturbation together realize a gate-tuned propagation geometry in which the excitation-induced response appears on both the gate-defined and mesa-defined trajectories.

One subtlety is that, in the absence of additional influences, the excitation at $V_\mathrm{c}=0$ would be expected to appear only along the mesa-defined path. However, a signal is also observed along the gate-defined path (Fig.~\ref{fig:mappings}(d)). This discrepancy is likely due to a modification of the surface potential caused by sidewall depletion and metal deposition, as discussed in Appendix B of the End Matter.

\begin{figure}
    \includegraphics[scale=1.0, pagebox=artbox, clip]{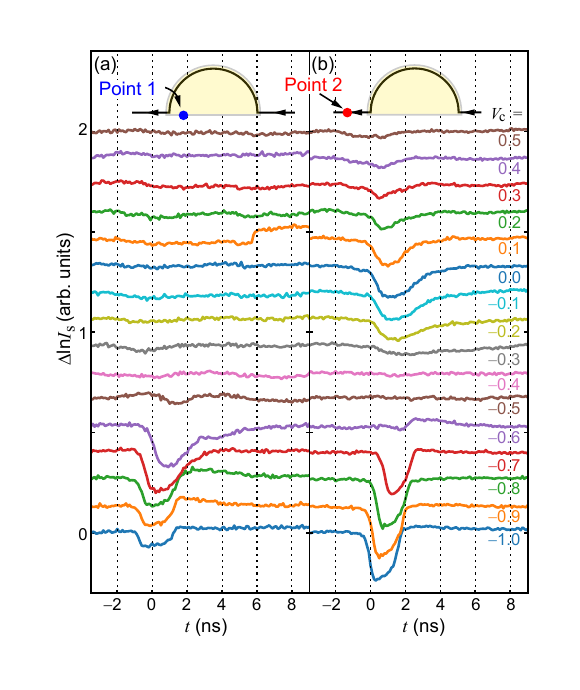}
    \centering
    \caption{Evolution of $\Delta \ln I_\mathrm{s}$ 
    as a function of $t$ measured at (a) Point 1 and (b) Point 2 (see Fig.~\ref{fig:device}). A negative-polarity pulse (levels $150~\mathrm{mV}$ to $-350~\mathrm{mV}$) is applied to the excitation gate. Each trace corresponds to a different control-gate voltage $V_\mathrm{c}$, as indicated on the right; traces are vertically offset for clarity. Each trace is collected with an exposure time of 20~s. Insets schematically indicate the measurement locations.}%
    \label{fig:time-resolve}%
\end{figure}

We next investigate edge-excitation dynamics on length scales exceeding $\sim 100~\mu$m. Because scanning the entire device was not feasible within experimental constraints, we focus on two fixed locations, Points~1 and~2 (Fig.~\ref{fig:device}). At these locations, we measure 
$\Delta \ln I_{\mathrm{s}}$
as a function of $t$ while varying  
$V_\mathrm{c}$ (Fig.~\ref{fig:time-resolve}).

For $V_\mathrm{c}=-1$~V, which is sufficient to deplete electrons beneath the control gate, clear dips appear in $\Delta \ln I_{\mathrm{s}}$   
at both Points~1 and 2 
(Fig.~\ref{fig:time-resolve}). The dip width at both points is $\sim 2$~ns, comparable to the width of the applied voltage pulse. Although the signal at Point~1 is weaker than that at Point~2, the waveform is similar, consistent with reduced optical throughput under the semi-transparent gate. These observations indicate that the excitation launched at the excitation gate propagates along a path comprising three stages: along the upstream ungated mesa edge, through the control-gate region, and along the downstream ungated mesa edge.

We now turn to the time delay of the observed waveforms. When the excitation follows the mesa-defined path ($V_\mathrm{c}=-0.2$~V to $0.4$~V), the trajectory is expected to shift inward as $V_\mathrm{c}$ decreases, which would shorten the path length and lead to an earlier arrival of the dip. Instead, Fig.~\ref{fig:time-resolve} shows that the dip is delayed by $\sim 1$~ns as $V_\mathrm{c}$ decreases. This indicates that the delay is not dominated by the path-length change but is primarily caused by a reduction in propagation speed as the confining potential becomes less steep when the edge moves away from the mesa, consistent with the local potential-gradient dependence of the propagation speed discussed above. The estimated velocity decrease is $\sim 5 \times 10^{4}$~m/s, consistent with Ref.~\cite{Kamata}. In contrast, when the excitation follows the gate-defined path ($V_\mathrm{c}=-1$~V to $-0.7$~V), decreasing $V_\mathrm{c}$ strengthens the confinement and advances the arrival by $\sim 1$~ns, consistent with an increased propagation speed.

We also comment on the dip amplitude. While it remains nearly constant at Point~2 for $V_\mathrm{c}\le -0.7$~V, it reaches a maximum at Point~1 near $V_\mathrm{c}=-0.7$~V and decreases for more negative voltages. The constant amplitude at Point~2 indicates that the propagation efficiency of the excitation itself is largely unchanged, whereas the reduced amplitude at Point~1 is consistent with a lateral shift of the gate-defined path away from the fixed optical focus at that measurement position.

\begin{figure}[!htbp]
    \centering
    \includegraphics[scale=1.0, pagebox=artbox, clip]{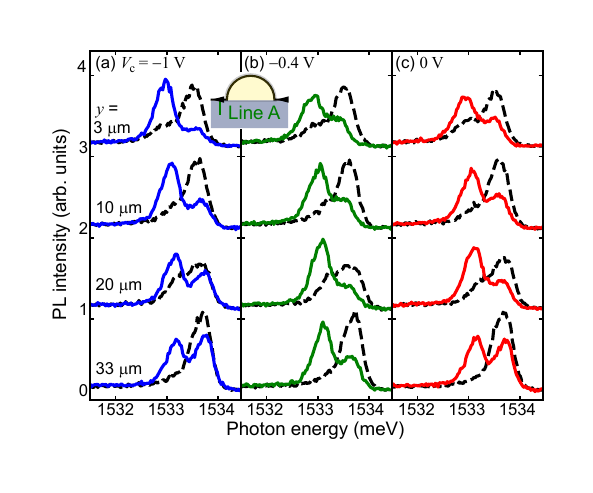}
    \caption{Micro-PL spectra acquired at fixed positions along Line~A, perpendicular to the mesa boundary defined as $y=0$, for different $V_\mathrm{c}$ with an exposure time of $15$~s. The spectra are shown for (a) $V_\mathrm{c}=-1$~V, (b) $-0.4$~V, and (c) $0$~V. The measurement positions are $y=3$, $10$, $20$, and $33~\mu$m, with positive $y$ taken toward the bulk side. The delay time is $t = -2.6$~ns. Colored solid curves show spectra measured with the excitation pulse, and black dotted curves show spectra measured without the pulse. 
    The excitation pulse amplitude used for these spectra is $\pm 200$~mV.}
    \label{fig:microPLspectra}%
\end{figure}

To examine how the excitation-induced PL spectral change depends on $V_\mathrm{c}$, we measured micro-PL spectra along Line~A at a fixed delay while varying $V_\mathrm{c}$ (Fig.~\ref{fig:microPLspectra}). The delay was set to $t=-2.6$~ns, corresponding to approximately $10$~ns after the preceding excitation pulse in the $13$-ns stroboscopic cycle; thus, Fig.~\ref{fig:microPLspectra} shows a delayed PL response rather than the immediate response to the voltage pulse. 
The full with- and without-pulse photon-energy--position maps are shown in Fig.~S3. Their comparison shows that the excitation voltage pulse modifies the local PL spectrum over the displayed gate-voltage range.
In particular, a clear spectral change remains visible at $V_\mathrm{c}=-0.4$~V, where a distinct dip is not resolved in the traces at Points~1 and~2. These pulse-induced spectral changes are unlikely to originate from simple heating, since comparable spectra measured upstream of the excitation gate do not show the same modification~\cite{jeong2026direct}. The PL changes extending tens of micrometers into the bulk side are consistent with the transverse near-field response and delayed secondary dynamics of an edge excitation analyzed in Ref.~\cite{jeong2026direct}.

Near the range in which the dip in Fig.~\ref{fig:time-resolve} disappears, from $V_\mathrm{c}=-0.5$ to $-0.4$~V, the dip systematically broadens.
Together with the clear persistence of the excitation-induced spectral change at $-0.4$~V seen in Fig.~\ref{fig:microPLspectra}, this broadening %evolution
motivates an interpretation in terms of velocity dispersion during propagation through the control-gated region. When the electrostatic potential in this region is tuned close to $E_\mathrm{F}$, the condition $U\approx E_\mathrm{F}$ can be satisfied over an extended area under the gate. In an ideal disorder-free system, this would correspond to a spatially delocalized edge position; in practice, however, the landscape is modulated by a weak disorder potential on the order of $0.1$~meV, commonly attributed to remote ionized donors in modulation-doped heterostructures \cite{eytan1998near,ilani2004microscopic,martin2004localization,hayakawa2013real}. In this situation, different portions of the response can sample different local confining-potential gradients and therefore acquire different propagation velocities, leading to temporal spreading of the excitation-induced waveform. Once the arrival-time distribution becomes comparable to the stroboscopic repetition period of $\sim 13$~ns, responses from successive excitation cycles overlap, reducing the visibility of a distinct dip associated with an individual pulse.

In summary, we have directly visualized delocalized edge excitations propagating along multiple paths in the fractional quantum Hall regime. By tuning the gate-induced potential landscape, we controlled the edge trajectory and observed propagation velocity dispersion consistent with multipath propagation shaped by disorder-induced potential variations.
These results identify a regime of edge dynamics in fractional quantum Hall systems characterized by long-range multipath propagation. This regime suggests a platform for analog quantum simulations of emergent spacetime geometries and may enable interference-based experiments in quantum Hall edge systems once phase coherence is established and verified. 
More broadly, a gate-defined quantum Hall edge in which an excitation can propagate through multiple controllable pathways provides a concrete analogue setting for ideas in which the effective spacetime geometry itself is not fixed to a single background. In this analogy, different edge-excitation pathways correspond to alternative effective geometrical backgrounds, offering a possible route toward experimental studies inspired by fluctuating or superposed spacetime geometries.

\begin{acknowledgments}
The authors thank T. Takayanagi and N. Shibata for fruitful discussions. This work was supported by a Grant-in-Aid for Scientific Research (Grant Nos.~21H05182, 21H05188, and 24H00399) from the Ministry of Education, Culture, Sports, Science, and Technology (MEXT), Japan. T.N. was supported by MEXT KAKENHI Grant Nos.~23K13094 and 24H00944 and by JST PRESTO Grant No.~JPMJPR2359. We are grateful to the YITP workshop YITP-T-25-01 held at YITP, Kyoto University, where part of this work was carried out.
\end{acknowledgments}

%\printbibliography
\bibliography{biblio}% Produces the bibliography via BibTeX.  <投稿するときはコメントにする。

\appendix
\section{End Matter}%\label{sec11}
\textit{Appendix A: Experimental methods}% \label{subsec:experimental}
---The stroboscopic PL measurements were performed with $\sim 1$-ps optical pulses from a mode-locked Ti:sapphire laser. 
The laser pulse train, with a repetition period \(T_{\mathrm{rep}}\approx 13\)~ns, was synchronized to the voltage pulses applied to the excitation gate. A pulse-pattern generator was used for the measurements in Fig.~\ref{fig:mappings}, Fig.~\ref{fig:time-resolve}, Fig.~S1, Fig.~S2(a), and Fig.~S2(b), whereas an arbitrary waveform generator was used for Fig.~\ref{fig:microPLspectra}, Fig.~S2(c), and Fig.~S3.
The beam was focused on the sample through an objective lens, giving a diffraction-limited spot size of 0.77~$\mu$m. The emitted PL was collected through the same objective lens and analyzed with a spectrometer equipped with a CCD detector. The focal position was controlled by piezoelectric stages. Each spectrum was accumulated over $N_{\mathrm{rep}}=T_{\mathrm{exp}}/T_{\mathrm{rep}}$ repetitions, with a total exposure time of $T_{\mathrm{exp}}=10$--$20$~s. The relative delay between the optical and electrical pulses was controlled electronically; the definitions of the delay origins used in Figs.~\ref{fig:mappings}--\ref{fig:microPLspectra} are given in the SM.\newline\hspace*{\parindent}
\textit{Appendix B: Effects of Metal Gate Deposition and Sidewall Depletion}% \label{effects_of_metal_gate} 
---Depositing a metal gate on the GaAs surface is known to modify the local electrostatic potential through the Schottky barrier determined by the metal work function and the pinning of surface states \cite{spicer1979new,freeouf1981schottky}, resulting in an offset relative to the bare GaAs surface. In addition, the etched mesa sidewall can induce lateral depletion and a corresponding local potential increase, particularly at low carrier density \cite{thornton1986one,simmons1988quantum,nieder1990one}. This sidewall-induced potential modulation can extend over several micrometers from the mesa edge \cite{grobecker2025enhancing}, facilitating depletion in the adjacent region.
\\[0.35\baselineskip]\indent
The combination of the metal-induced surface-potential offset and sidewall depletion can therefore produce a non-negligible built-in electrostatic modulation even at $V_\mathrm{c}=0$, which can shift the effective edge position toward the gate boundary and allow edge excitations to propagate along the gate-defined path. Consistent with this interpretation, a positive control-gate voltage of approximately $+0.3$~V is required to compensate this built-in offset and confine the edge excitation strictly to the mesa-defined path.

\clearpage
\onecolumngrid
\input{SM_20260609_body}
\end{document}

%% file: SM_20260609_body.tex
% This file is intended to be included from EdgeControl_arxiv.tex using \input.
% Do not add \documentclass, \usepackage, \begin{document}, \maketitle, or \end{document} here.

\clearpage
\setcounter{section}{0}
\setcounter{figure}{0}
\setcounter{equation}{0}
\renewcommand{\figurename}{FIG.}
\renewcommand{\thefigure}{S\arabic{figure}}
\renewcommand{\theequation}{S\arabic{equation}}

\begin{center}
{\Large\bfseries Supplemental Material}\\[1.0em]
{\normalsize
Yunhyeon Jeong, Akinori Kamiyama, John N. Moore, Takaaki Mano, Yuuki Sugiyama,\\
Tokiro Numasawa, Ken-ichi Sasaki, Masahiro Hotta, Go Yusa
}
\end{center}

\vspace{1.0em}

\section{Definition of the logarithmic change in the singlet PL intensity}
 
\begin{figure}[htbp]
  \centering
  \includegraphics[width=0.9\linewidth]{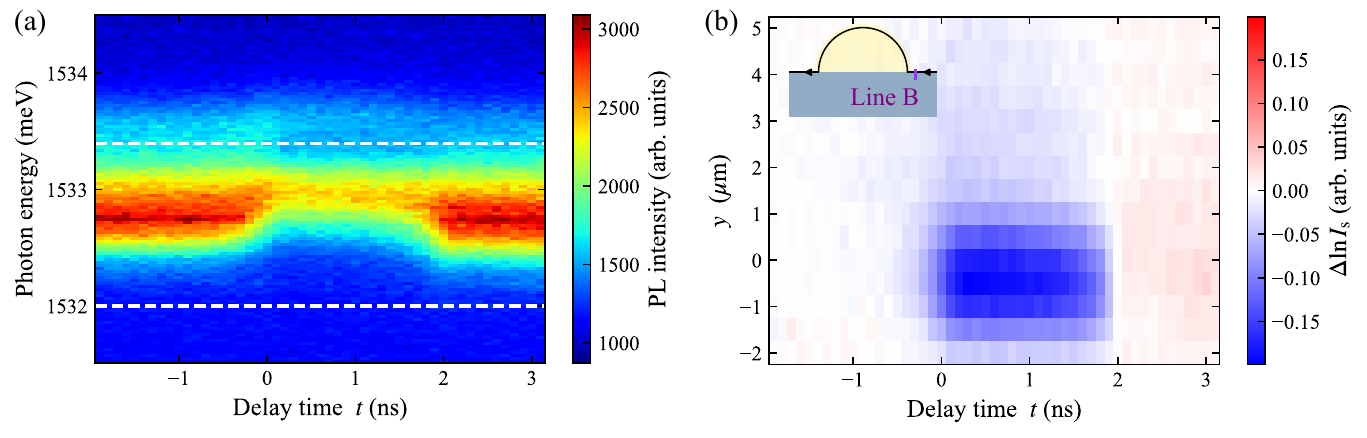}
  \caption{Definition of the logarithmic change in the singlet PL intensity.
  (a) Representative time-resolved micro-PL spectrum measured at the sample-edge position ($y = 0$) on Line~B. The color scale shows the PL intensity on a linear scale. The white dashed lines indicate the photon-energy window from $1532.0$ to $1533.4$~meV used to define $I_\mathrm{s}$.
  (b) Logarithmic change in singlet PL intensity $\Delta \ln I_\mathrm{s}$ measured along Line~B as a function of transverse position $y$ and delay time $t$. The inset schematically shows the measurement position of Line~B relative to the control gate. Line~B is located approximately $7~\mu$m on the $-x$ side of the right-hand edge of the control gate, and $y$ was scanned from $-2$ to $5~\mu$m.}
  \label{fig:sm-input-excitation}
\end{figure}

Fig.~\ref{fig:sm-input-excitation} explains how the logarithmic change in the singlet PL intensity used throughout this work is obtained from time-resolved micro-PL spectra. Fig.~\ref{fig:sm-input-excitation}(a) shows a representative time-resolved micro-PL spectrum measured at the sample-edge
position ($y = 0$) on Line~B. The measurement position of Line~B is shown schematically in the inset of Fig.~\ref{fig:sm-input-excitation}(b). The delay coordinate $t$ used here is defined using the origin shown in Fig.~\ref{fig:sm-time-origin}(a). For each delay time, the integrated singlet PL intensity $I_\mathrm{s}$ is defined by integrating the PL intensity over the photon-energy window from $1532.0$ to $1533.4$~meV, indicated by the white dashed lines.
 
The logarithmic change in singlet PL intensity is then evaluated as
\begin{equation}
  \Delta \ln I_\mathrm{s}(t) \equiv \ln \left[ I_\mathrm{s}(t) / I_\mathrm{s}^0 \right],
  \label{eq:sm-singlet-pl-change}
\end{equation}
where $I_\mathrm{s}^0$ is the reference integrated singlet PL intensity obtained from the same energy-integration window. The reference intensity $I_\mathrm{s}^0$ is defined separately for each time-resolved data set. Unless otherwise noted, it is taken from the first measured delay point of the corresponding time trace, which lies before the arrival of the edge excitation. For spatially resolved measurements, this procedure is applied independently at each measured position, so that $I_\mathrm{s}^0$ is defined locally for each spatial point. In particular, for the spatial maps in Fig.~2(b)--(f), $\Delta \ln I_\mathrm{s}$ was evaluated as $\ln [I_\mathrm{s}(t = 0.6~\mathrm{ns}) / I_\mathrm{s}(t = -1.9~\mathrm{ns})]$ at each spatial point.
 
Fig.~\ref{fig:sm-input-excitation}(b) shows the resulting $\Delta \ln I_\mathrm{s}$ measured along Line B as a function of the transverse position $y$ and delay time $t$. Negative values of $\Delta \ln I_\mathrm{s}$ correspond to suppression of the integrated singlet PL intensity relative to the pre-arrival reference. This normalized quantity is used for the spatial maps in Fig.~2(b)--(f), the time traces in Fig.~3, and the delay-time definitions described below.
 
\section{Definition of delay origins}
The delay time in this stroboscopic measurement is defined as the relative delay between the optical pulse from the Ti-sapphire laser and the voltage pulse applied to the excitation gate. Since the measurement is performed periodically, the delay is defined modulo the repetition period. For each measurement configuration, the delay origin was chosen from a representative excitation-induced response, as shown in Fig.~S2. After this origin shift, the delay is denoted simply by $t$ throughout the manuscript and the Supplemental Material.

The origin shown in Fig.~S2(a) is used for Figs.~2(b)--(f) and Fig.~\ref{fig:sm-input-excitation}. The origin shown in Fig.~S2(b) is used for Fig.~3. The origin shown in Fig.~S2(c) is used for Fig.~4 and Fig.~S3.

%In this work, we use three delay-time coordinates, $t_1$, $t_2$, and $t_3$, depending on the measurement configuration. The coordinate $t_1$ is used for Figs.~2(b)--(f) of the manuscript and Fig.~\ref{fig:sm-input-excitation}; $t_2$ is used for Fig.~3 of the manuscript; and $t_3$ is used for Fig.~4 of the manuscript and Fig.~S3.
 
The delay %time
origins were defined using the excitation-induced logarithmic change in the singlet PL intensity $\Delta \ln I_\mathrm{s}$ defined in Sec.~1. For each measurement configuration %delay-time coordinate $t_i$,
the delay origin %$t_{i,0}$ 
was chosen as the delay at which the numerical derivative of $\Delta \ln I_\mathrm{s}$ %$d (\Delta \ln I_\mathrm{s})/dt_i$
took its minimum value. This corresponds to the steepest falling point of $\Delta \ln I_\mathrm{s}$ induced by the arrival of the edge excitation. This procedure follows the same convention as that used in Ref.~\cite{jeong2026direct}, except that the present analysis uses the integrated singlet PL intensity defined in Sec.~1 rather than the fitted singlet PL amplitude. This difference does not affect the role of the delay %time
origin, because the origin is used only as an operational timing within each measurement configuration.
 
\begin{figure}[htbp]
  \centering
  \includegraphics[width=0.99\linewidth]{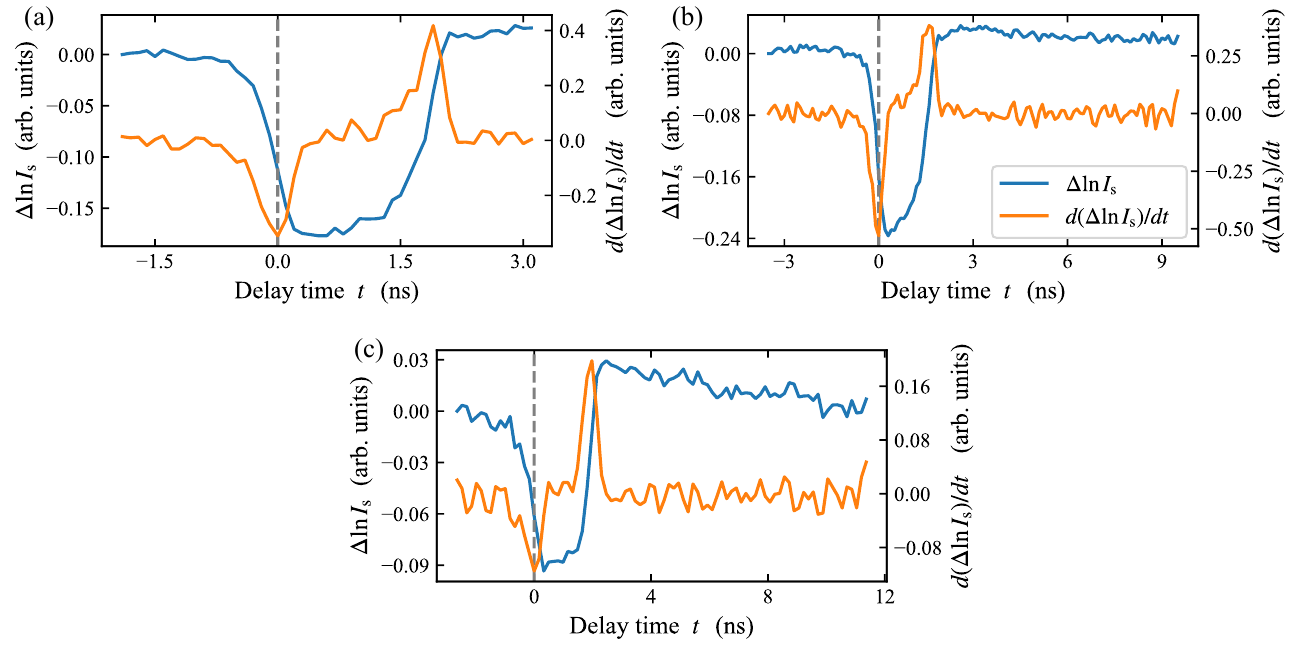}
  \caption{Definition of the delay origins used for different measurement configurations. The blue curves show the logarithmic change in singlet PL intensity $\Delta \ln I_\mathrm{s}$, and the orange curves show its numerical derivative with respect to the delay. The gray dashed lines indicate the defined origins. (a) Definition of $t_{1,0}$ from the response measured at $y = 0$ on Line B. (b) Definition of $t_{2,0}$ from the $V_c = -1.0$ V trace measured at Point 2 in Fig.~3(b) of the manuscript. (c) Definition of $t_{3,0}$ from the response measured at $y = 0$ on Line A.}
  \label{fig:sm-time-origin}
\end{figure}

Fig.~\ref{fig:sm-time-origin} shows the definitions of the three delay-time origins. Fig.~\ref{fig:sm-time-origin}(a) shows $\Delta \ln I_\mathrm{s}$ measured at $y = 0$ on Line B under the same pulse condition as that used for Figs.~2(b)--(f) of the manuscript. The gray dashed line indicates the origin $t_{1,0} = 0$ ns. Fig.~\ref{fig:sm-time-origin}(b) shows $\Delta \ln I_\mathrm{s}$ extracted from the $V_\mathrm{c} = -1.0$ V trace at Point 2 in Fig.~3(b) of the manuscript, measured under the same pulse condition as that used for Fig.~3. The gray dashed line indicates the origin $t_{2,0} = 0$ ns. Fig.~\ref{fig:sm-time-origin}(c) shows $\Delta \ln I_\mathrm{s}$ measured at $y = 0$ on Line A under the same pulse condition as that used for Fig.~4 of the manuscript. The gray dashed line indicates the origin $t_{3,0} = 0$ ns.
 
Unless otherwise noted, the delay is denoted simply by $t$ after its origin has been specified.
 
\section{Estimation of Electrochemical Potential Shift} \label{subsec:estimation}
 
The change in the 2DEG density induced by the control-gate voltage can be estimated using a parallel-plate capacitor model, $e\,\Delta n_e = C\,\Delta V$, where $C$ is the capacitance per unit area between the control gate and the 2DEG. Using the device geometry, we obtain $\Delta n_e/\Delta V = 3.9\times10^{11}~\mathrm{cm^{-2}/V}$. To convert this density change into an energy scale, we use the zero-field 2D relation between carrier density and chemical potential, $\Delta \mu \simeq (\pi\hbar^2/m^\ast)\Delta n_e$, where $m^\ast=0.067\,m_e$ is the GaAs effective mass. This yields an estimated electrochemical-potential shift of $\Delta \mu \sim 4$~meV for $\Delta V \sim 0.3$~V, corresponding to the voltage range over which two paths coexist in Figs.~2(d) and 2(e). For the excitation-gate pulse used in Fig.~2(b)--(f), the voltage excursion of $0.2~\mathrm{V_{pp}}$, equivalently denoted as a $\pm 0.1~\mathrm{V}$ pulse, gives an energy scale of $\sim 3~\mathrm{meV}$. This value is used only as an order-of-magnitude scale for comparing the transient excitation-gate perturbation with the control-gate voltage range over which the two paths coexist. These estimates are intended as order-of-magnitude energy scales; in the quantum Hall regime the relation between $\Delta n_e$ and $\Delta\mu$ can be modified by Landau quantization and screening, but the comparison of the relevant voltage scales remains valid.
 
\section{Micro-PL spectra along Line A for different control-gate voltages}
 
\begin{figure}[htbp]
  \centering
  \includegraphics[scale=0.8]{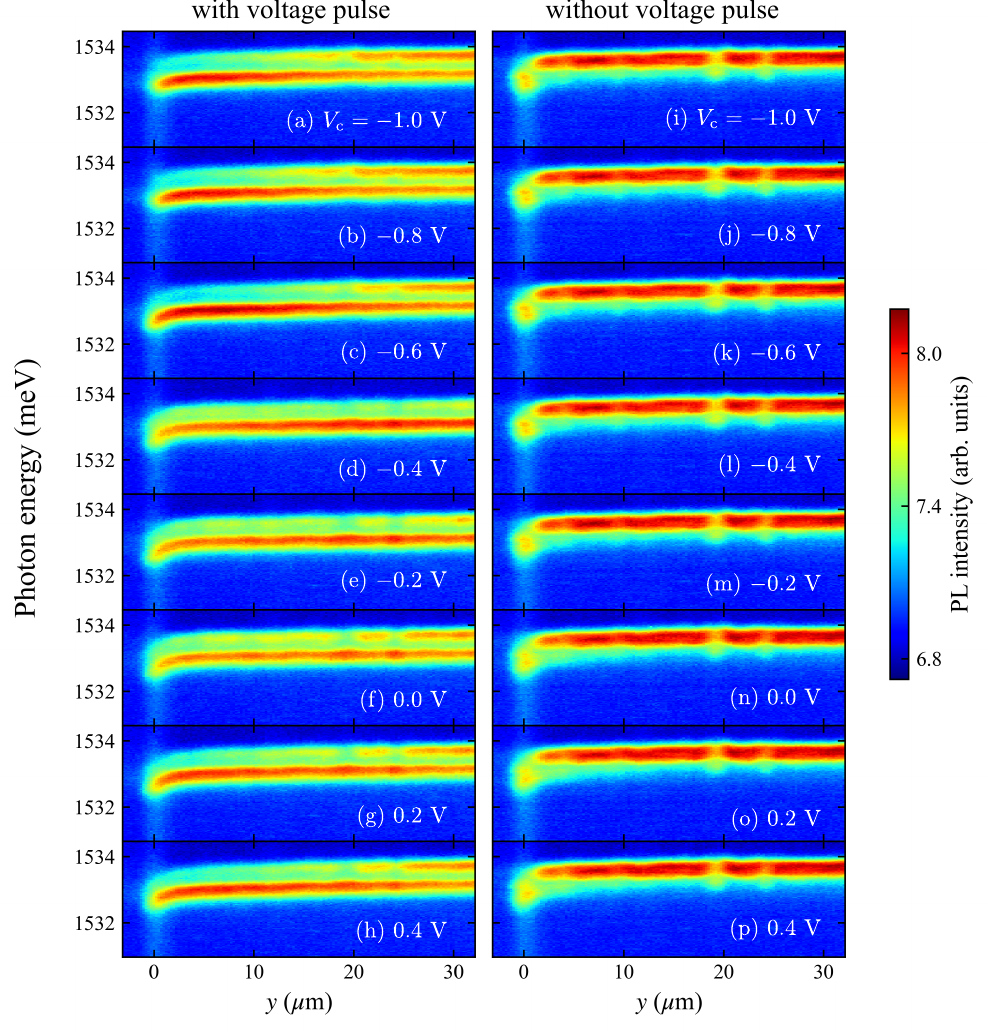}
  \caption{Color maps of micro-PL spectra measured as functions of position $y$ along Line A for several values of $V_c$ ranging from $-1$ to $0.4$~V. Panels (a)--(h) show spectra measured with the excitation voltage pulse applied under the same experimental conditions as those used for Fig.~4 of the manuscript, with an excitation-pulse amplitude of $\pm 200$~mV. The delay time was fixed at $t = -2.6$~ns. Panels (i)--(p) show the corresponding spectra measured without applying the excitation voltage pulse. The PL intensity is plotted on a logarithmic color scale.}
  \label{fig:sm-micro-PL-lineA}
\end{figure}

Fig.~\ref{fig:sm-micro-PL-lineA} shows color maps of micro-PL spectra acquired along Line A for several values of $V_c$, plotted as functions of photon energy and position $y$. Panels (a)--(h) were measured with the excitation voltage pulse applied, under the same experimental conditions as those used for Fig.~4 of the manuscript. The delay time was fixed at $t = -2.6$~ns, and the representative spectra shown in Fig.~4 of the manuscript were extracted from this data set.
 
Panels (i)--(p) show the corresponding spectra measured without applying the excitation voltage pulse. The comparison between the with-pulse and without-pulse data shows the excitation-induced modification of the local PL spectrum along Line A. The localized spatial intensity modulations, such as the local minima around $y \simeq 20$ and $24$~$\mu$m, indicate spatial inhomogeneity of the PL response along the scan line.
 
